\documentclass[conference]{IEEEtran}
\IEEEoverridecommandlockouts
\usepackage{cite}
\usepackage{amsmath,amssymb,amsfonts}
\usepackage{algorithmic}
\usepackage{tabularx}
\usepackage{comment}
\usepackage{amsmath}
\usepackage{relsize}
\usepackage{graphicx}
\usepackage{textcomp}
\usepackage[table,xcdraw]{xcolor}
\usepackage{xcolor}
\def\BibTeX{{\rm B\kern-.05em{\sc i\kern-.025em b}\kern-.08em
    T\kern-.1667em\lower.7ex\hbox{E}\kern-.125emX}}
\begin{document}

\title{Using Transition Learning to Enhance Mobile-Controlled Handoff In Decentralized Future Networks\\
\thanks{This work was partially supported by the Spanish and Catalan Governments through the project "Plan Nacional": AEI/FEDER TEC2016-79510 "Redes Con Celdas Densas y Masivas" and the SGR2017-2019.}
}

\author
{\IEEEauthorblockN{1\textsuperscript{st} Steven Platt}
\IEEEauthorblockA{
\textit{DTIC, Universitat Pompeu Fabra}\\
Barcelona, Spain \\
steven@ieee.org}

\and
\IEEEauthorblockN{2\textsuperscript{nd} Berkay Demirel}
\IEEEauthorblockA{
\textit{DTIC, Universitat Pompeu Fabra}\\
Barcelona, Spain \\
berkay.demirel@upf.edu}

\and
\IEEEauthorblockN{3\textsuperscript{rd} Miquel Oliver}
\IEEEauthorblockA{
\textit{DTIC, Universitat Pompeu Fabra}\\
Barcelona, Spain \\
miquel.oliver@upf.edu}

}

\maketitle

\begin{abstract}
Traditionally, resource management and capacity allocation has been controlled network-side in cellular deployment. As autonomicity has been added to network design, machine learning technologies have largely followed this paradigm, benefiting from the higher compute capacity and informational context available at the network core. However, when these network services are disaggregated or decentralized, models that rely on assumed levels of network or information availability may no longer function reliably. This paper presents an inverted view of the resource management paradigm; one in which the client device executes a learning algorithm and manages its own mobility under a scenario where the networks and their corresponding data underneath are not being centrally managed. 
\end{abstract}


\section{Introduction}
Network softwarization in 5G has allowed unprecedented flexibility in how cellular services are configured and delivered. Moving from traditional MVNO agreements and overlay networks existing with 4G, to enabling every function of the network with the ability to be virtualized and made dynamic in 5G and beyond deployment. As previously seen in cloud computing, this rapid advance of software has encouraged a decoupling of hardware from software to the extent that slower moving hardware generations are made general purpose and are able to accommodate increasing heterogeneity of software and services sitting on top \cite{gcomm-b3}. A potential of such decoupling is that in the long term, network infrastructure can be fully disaggregated to the extent that it becomes possible to stitch together wholly new formats of service from multiple Amazon Web Services for \textit{5G ...6G ...and beyond}. 

SDN (Software Defined Networking) technologies which previously allowed decoupling of data and control planes for backbone network flows are increasingly being adapted for wireless. These include recent research for adversarial dynamic spectrum access and software radio to enable infrastructure slicing through to the radio access edge \cite{gcomm-b3}. In tandem with these advances, standardization activities including the ETSI GANA (Generic Autonomous Networking Architecture) now provide a reusable model for the separation of higher-level resource orchestration (the cellular control plane) and the dynamic and software driven heterogeneous infrastructures delivering services underneath \cite{gcomm-b4}\cite{gcomm-b20}. Extending further from a general separation of data and control planes, recent research and commercial offerings increasingly are pursuing a goal of delivering infrastructures and services piecemeal or decentralizing and abstracting away service providers entirely. These range from a basic expansion of classic MVNO models such as Google Fi \cite{gcomm-b5} and HMD Connect \cite{gcomm-b6}; to dynamic and API consumable wireless services from vendors such as Twilio \cite{gcomm-b7} and Telnyx \cite{gcomm-b8}; and finally a full decentralization of wireless network functions using blockchain technologies \cite{gcomm-b9}\cite{gcomm-b29}\cite{gcomm-b30}. 

One difficulty in realizing full decentralization however is classic resource management structures which retains global visibility at the base station and cellular core, paired with a subordinate UE (User Equipment) device at the edge. Taking this as a starting point; one question raised is what becomes of the UE device at the network edge and how are network services consumed in the absence of classic network control. With significantly less environmental context available at the UE, addressing device control under this general lack of data requires further research. This paper investigates an enhancement of existing mobile-controlled handoff capabilities by doing all learning \textit{on-device}" using the existing mechanic of measuring RSSI (Received Signal Strength Indicator). The remainder of the paper is split into four parts. First we provide a background into the existing mechanics of cellular mobility, recent research into cellular network decentralization, and potential machine learning methods that may be considered as alternatives to the approach detailed in this paper. After this we present our design of a "transition learning" algorithm in section three, followed by our simulation results in section four. The paper concludes with a discussion of results and identification of paths for future research in section five. 


\section{Background and Related Research}

The following section provides additional background and related research to highlight the gap and contribution made by the transition learning algorithm being presented in this research. This section covers cellular mobility management, network decentralization, and machine learning applications and limitations. 

\subsection{Cellular Mobility Management}

Across network generations and vendor configurations, cellular mobility can inherit a broad range of architectures. At a high level, these can be organized into three categories: network controlled handoff, mobile assisted handoff, and mobile controlled handoff \cite{gcomm-b14}.

\subsubsection{Network Controlled Handoff} As the most centralized approach, network control handoff places all knowledge and mobility control with network base stations. This approach is largely a carryover of the earliest network design in which UE devices lacked the sensors and compute resources to participate in mobility coordination. In this model, mobility decisions not taken at the network edge can add a non-trivial amount of signaling latency if they include a remote or regional network core.


\subsubsection{Mobile Assisted Handoff} UE devices participating in mobile assisted handoff are able to report on-device sensor readings, specifically RSSI which is calculated from the RSRP (received signal received power) and RSRQ (reference signal received quality)\cite{gcomm-b12}. With this data updating periodically, the core network is able to balance the state of the UE device, with its global visibility of the wider capability and status of the network, including total device density, its own backhaul capacity, and the specific commitments and priorities tied to all other services operating from a given base station ahead of making a mobility decision. 

\subsubsection{Mobile Controlled Handoff} 
\begin{figure}
\centering
\includegraphics[width=0.7\linewidth]{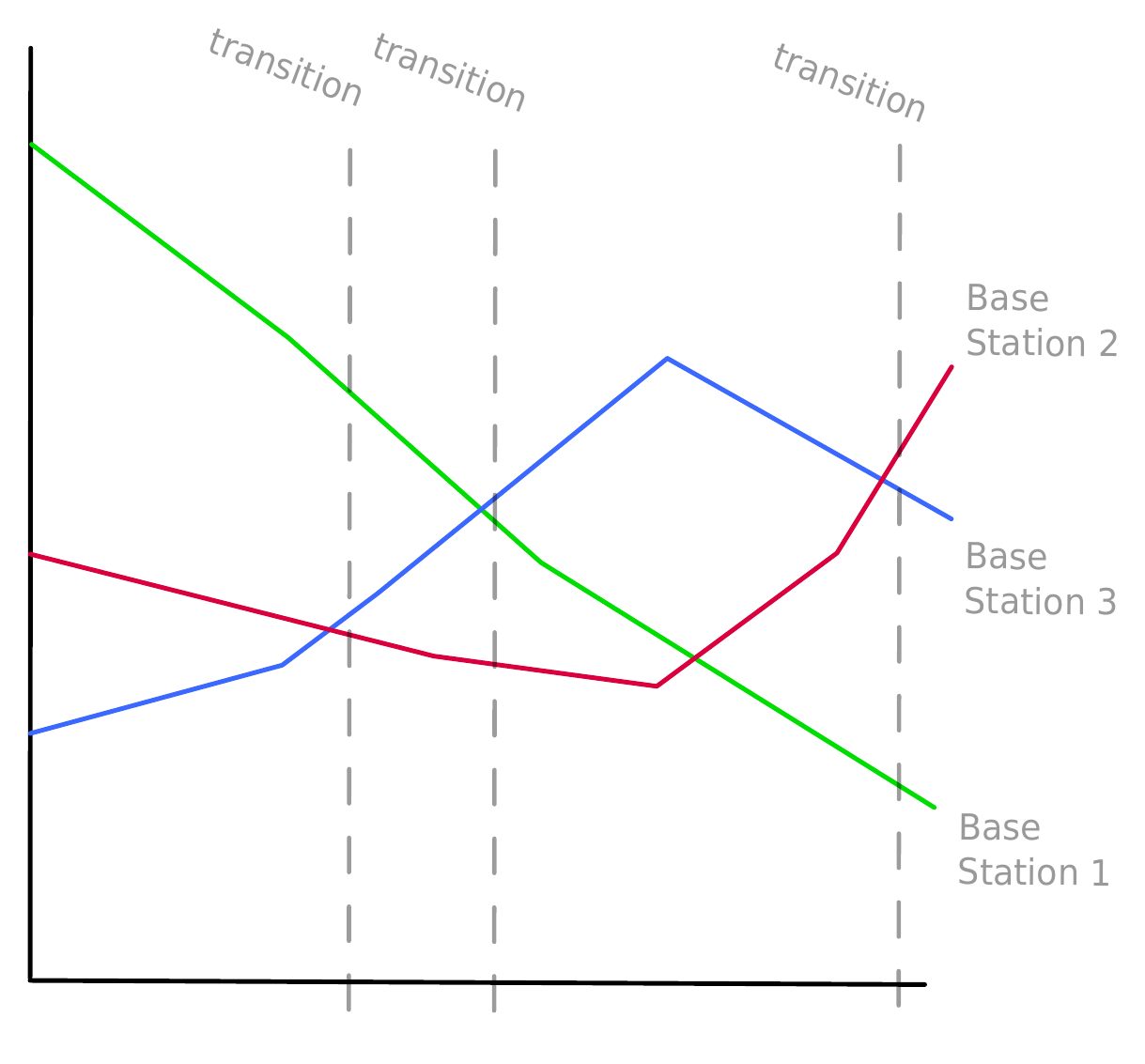}
\caption{Example transitions based on RSSI received at the UE.}
\label{classic_mobility}
\end{figure}

Allowing the UE to handle handoff decisions reduces handoff times compared to the previously mentioned methods \cite{gcomm-b2}. In this model, the UE monitors the measured RSSI values of pilot channels signals received from surrounding base stations and initiates a handoff when certain conditions are met, such as when the RSSI from a connected base station is no longer the highest and drops below a defined threshold with additional padding to limit hysteresis (Fig. \ref{classic_mobility}) \cite{gcomm-b1}\cite{gcomm-b13}. The research and transition learning algorithm presented in this paper are targeted at extending the capability of this type of handoff operation.

\subsection{Blockchain Network Decentralization}

\begin{figure}
\centering
\includegraphics[width=\linewidth]{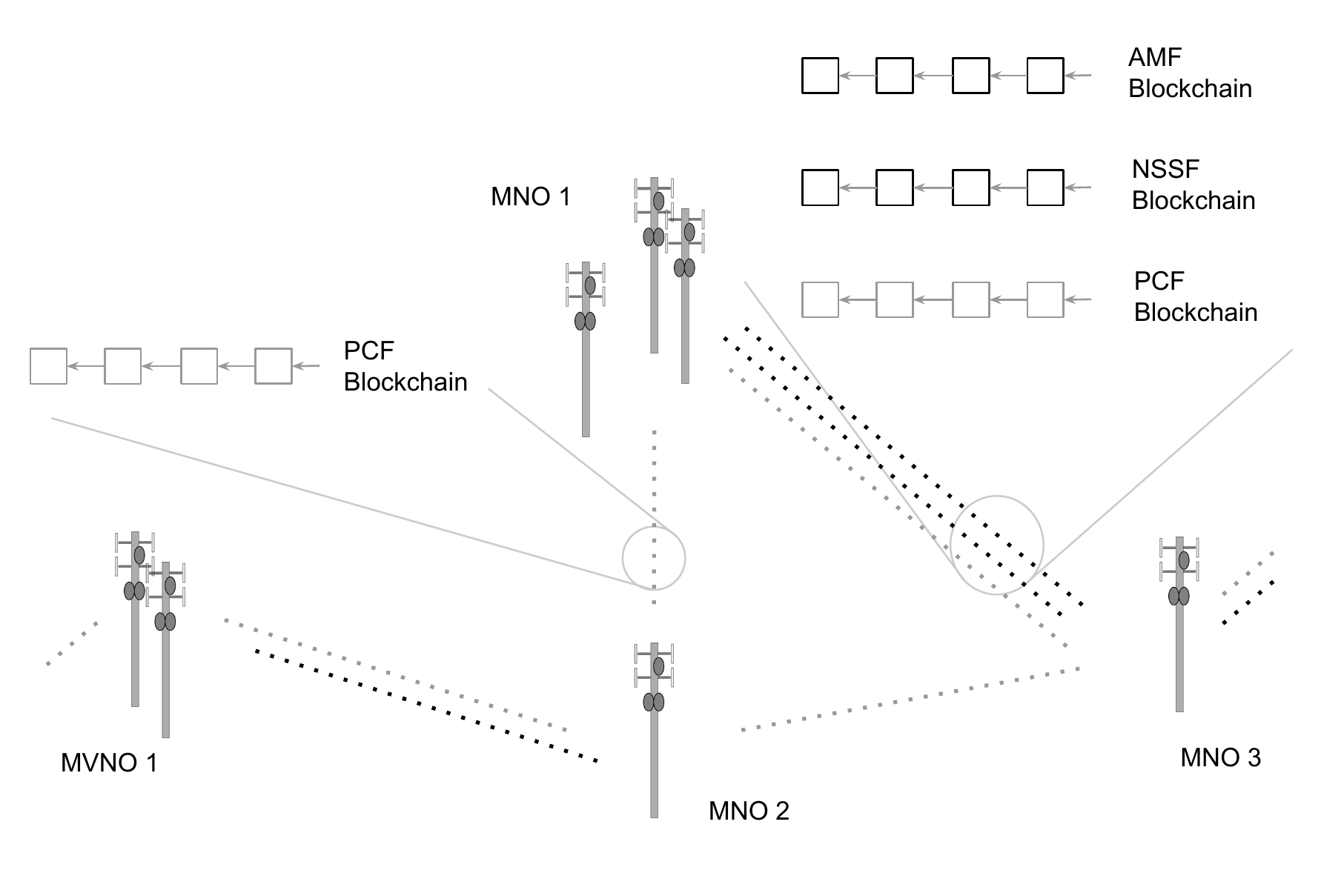}
\caption{Example networks operating individually decentralized network functions \cite{gcomm-b9}.}
\label{network_blockchain}
\end{figure} 

Blockchain at its lowest level is a forward hash-linked data structure. Data stored in "blocks" are hashed and this hash is carried forward and added to new blocks which themselves are then hashed. By including the hash from previous blocks, the data in total becomes cryptographically linked, forming a "chain" \cite{gcomm-b36}. Blockchain technology encompasses an entire category of implementations supporting combinations of cryptocurrency and contracts logic \cite{gcomm-b32} or isolated to be used only as database storage \cite{gcomm-b9}. 

In network implementation, blockchain has been pursued to allow a broad decentralization of network infrastructures and services. Examples include applications of network access control \cite{gcomm-b29}\cite{gcomm-b35}, spectrum access auctions \cite{gcomm-b34}, and the general use of blockchain technology as an agnostic storage layer used by network functions (fig. \ref{network_blockchain}) \cite{gcomm-b9}. The latter is significant because it is intended to be generalizable and allow broad decentralization of any network service which is built atop 5G VNF's (Virtual Network Functions).

While this paper is not an investigation of blockchain technology itself, it is an important context to highlight as the experiment presented assumes a network environment where the infrastructures are not part of a unitary carrier deployment, but are instead independent with the only commonality being the UE which has access across them. This context is most similar to emergency calling or WPS (Wireless Priority Service) in which a UE, even while not having active carrier subscription, must be permitted access to available networks when placing emergency calls. To the author's knowledge, there is no deployed available equivalent to WPS for data access \cite{gcomm-b33}. The presented research extends the current body of knowledge in this direction.

\subsection{Learning Applications and Limitations}

Machine learning is a very active path of investigation for enabling autonomy in decision making. Machine learning approaches can be classified into three broad categories depending on the type of feedback signal available to the learning system: supervised learning, unsupervised learning, and reinforcement learning. This section provides a summary of these three as well as a fourth, more narrow subcategory chosen for the experiment in this paper. 

\subsubsection{Supervised Learning} Supervised learning models learn to generalize the input-output mappings presented to it by a “supervisor” signal in the form of labeled data. The use of labeled data to train and predict new data points gives precise control of what the model learns through the curation of the labeled dataset. Training supervised learning systems with high quality data that is representative of the ground truth can lead to high levels of accuracy in unseen data points. This level of precise control of what the model learns and dependence on labeled data points is also a drawback of supervised learning systems, as they require both a large and varied amount of representative data to be able to generalize well. Supervised learning methods are less common in cellular deployment, but have been employed for mobile edge computing (MEC) and QoS policy control operations taking place at the less resource-constrained network core \cite{gcomm-b21}\cite{gcomm-b22}.

\subsubsection{Unsupervised Learning} In cases where labeled data is difficult to acquire or outright not available, unsupervised learning approaches can be used to uncover the underlying structures in data. These approaches trade a level of control on what the model learns for the ability to learn underlying structures and make predictions without knowing the ground truth in the form of labeled data. Beyond also requiring a large and varied dataset, a second drawback specific to unsupervised learning is the difficulty in assessing the accuracy of these models derived from unlabeled data without human validation. Human effort is back-loaded with unsupervised learning, compared to supervised learning where most human effort is front-loaded through the labeling of datasets to ensure they represent a ground truth. In 5G and beyond contexts, the unstructured format of unsupervised data learning has been has often been paired with network stream data and monitoring systems for retroactive self-diagnosis rather than autonomous actuation of cellular resources due to the mentioned lack of control over \textit{what} is learned \cite{gcomm-b23}\cite{gcomm-b24}.

\subsubsection{Reinforcement Learning} Reinforcement learning models learn to map actions to situations in order to maximize a designated reward. A reinforcement learning approach differs from the previous two approaches in multiple ways. First of all, instead of learning from a large dataset, reinforcement learning agents interact with an environment to gather data points and learn how to maximize a reward signal. As the reinforcement learning agent is naive about the environment, it is required to explore the environment as well as exploit any potential source of rewards. This ongoing dilemma between exploration and exploitation means reinforcement learning agents require a high level of interaction. As a result of this ongoing interaction, it is much more adaptable with a minimum need for retraining as it can slowly adapt to changes with each new interaction while expiring old data, allowing it to be relatively storage efficient compared to supervised and unsupervised machine learning models. This structure and behavior make reinforcement learning a better suited candidate for cellular application and operations managed by the UE. Reinforcement learning models are commonly used in wireless for decision making under unknown network conditions and contexts involving resource competition or opportunistic access \cite{gcomm-b25}\cite{gcomm-b26}\cite{gcomm-b27}\cite{gcomm-b28}.

\subsubsection{Markov Chains and Transition Learning}

\begin{figure}
\centering
\includegraphics[width=\linewidth]{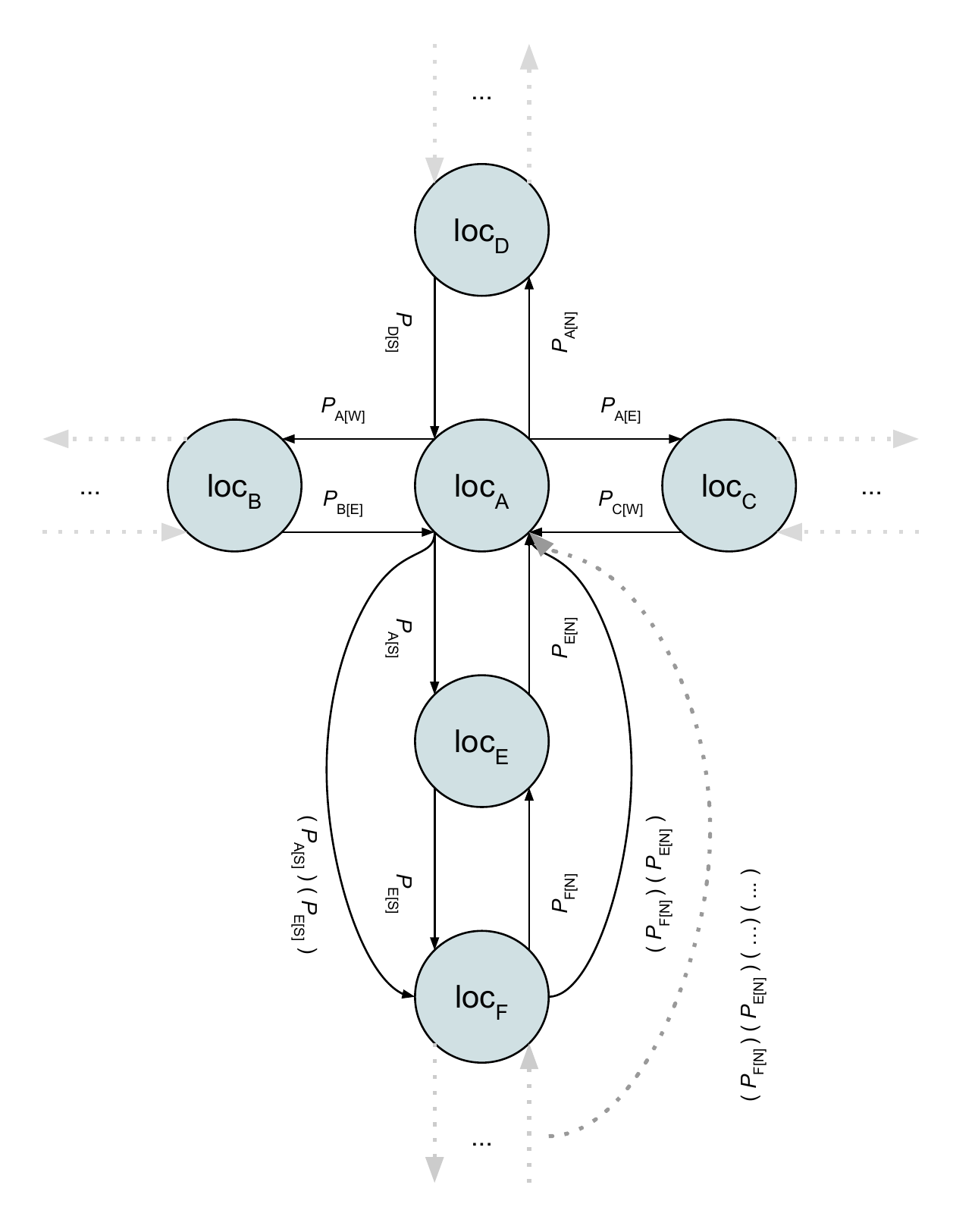}
\caption{2-D Markov Chain model with North [N], South [S], East [E], and West [W] transitions.}
\label{markov_chain}
\end{figure}

Markov Chains are a method of representing the probabilities of moving from one state to another. This movement is referred to as a \textit{transition}. By design Markov Chains and Markov processes are intended to model an expected outcome based only on a current state and are considered \textit{memoryless} \ref{markov_chain}. Markov chains are often used to model processes that are stochastic and where past history has decaying or no value over time such as in wireless networks \cite{gcomm-b15}\cite{gcomm-b16}. In cases where additional context can be gained from previous states, these Markov transitions can be saved for further processing in the form of Transition Learning. Data produced during state transitions in cellular networks has also been used as the training set for the previously mentioned formats of machine learning \cite{gcomm-b17}\cite{gcomm-b18}. This paper applies transition learning in isolation, rather than within a large learning algorithm. To the authors knowledge, transition learning has not previously been investigated in isolation as a solution to extend the capabilities of mobile-controlled handoff.  

Although a large block of learning algorithms fall into one of these broad categories, they should be understood more as general areas, and less as strict separations as there are exceptions existing which do not map cleanly into a single category as seen with methods such as meta-learning which can provide a cross-category aggregated result \cite{gcomm-b19}.

\section{System Design}

In this section we aim to implement an algorithm that extends the capabilities of the RSSI data already existing at the UE to determine if this minimal amount of data can be used to help a given UE take higher performing base station associations under a scenario of mobile-controlled handoff. To do this, we create an algorithm where a UE can store and take decisions informed by a compact history of prior state transitions combined with the performance outcome it received (fig. \ref{algorithm}. The following section details the transition learning algorithm and setup of the simulation environment. 

\subsubsection{Base Station Association} In the experiment, it is assumed that the UE has access and a policy giving equal preference to all base stations in the environment. In order to represent a traditional preferred roaming list, the UE constantly monitors the 3 closest base stations. The UE is configured to always associate with the closest base station of the three, mimicking default RSSI association behavior. 

\subsubsection{Base Station Allocation} Because real world cellular performance is a temporal mix of frequency band, resource block allocation, signal interference, backhaul load and further factors - the experiment abstracts these and defines an "allocation" value to be used as a proxy representing composite performance measured at the UE. Further, the experiment treats the base station allocations as uniform with an isotropic radiation pattern in free space. Allocation values of 5 and 7 were used to present a scenario of significant $allocation\ \Delta$ (fig. \ref{grid_world}).

\subsubsection{Defining Transitions}

To define transitions, the UE begins in some $state$ where it checks for the 3 base stations with the highest RSSI defined by their physical proximity (\ref{ranking}). After completing a random walk, the UE checks whether the rank order of these strongest 3 signals is changed. If it is not changed, the UE does not have a new state and does not evaluate any mobility action. If the UE detects a change in the rank order \emph{and} the strongest signal is also changed (\ref{new_state}); the UE understands this as a new $state'$. From here the UE takes the default action of connecting to the base station with the strongest signal and calculates the difference in the allocation it received from moving to the new $state'$ as an $allocation\ \Delta$. This beginning $state$, final $state'$ and $allocation\ \Delta$ are stored as $transition_n$ (\ref{transition}). This $transition_n$ is the only value the UE retains in memory.

\begin{equation}
\begin{array}{c}
state\ =\\ 
\begin{bmatrix}base\ station\ rank_1, \\
base\ station\ rank_2, \\
base\ station\ rank_3\end{bmatrix}
\end{array}
\label{ranking}
\end{equation}
\vspace{6pt} 

\begin{equation}
\begin{array}{c}
state(base\ station\ rank_1)\ \not= \\
state'(base\ station\ rank_1)
\end{array}
\label{new_state}
\end{equation}
\vspace{6pt} 

\begin{equation}
\begin{array}{c}
transition_n =\\ 
("state", "state'", "allocation\ \Delta")
\end{array}
\label{transition}
\end{equation}
\vspace{6pt}

\subsubsection{Transition Learning} 

\begin{figure}
\centering
\includegraphics[width=\linewidth]{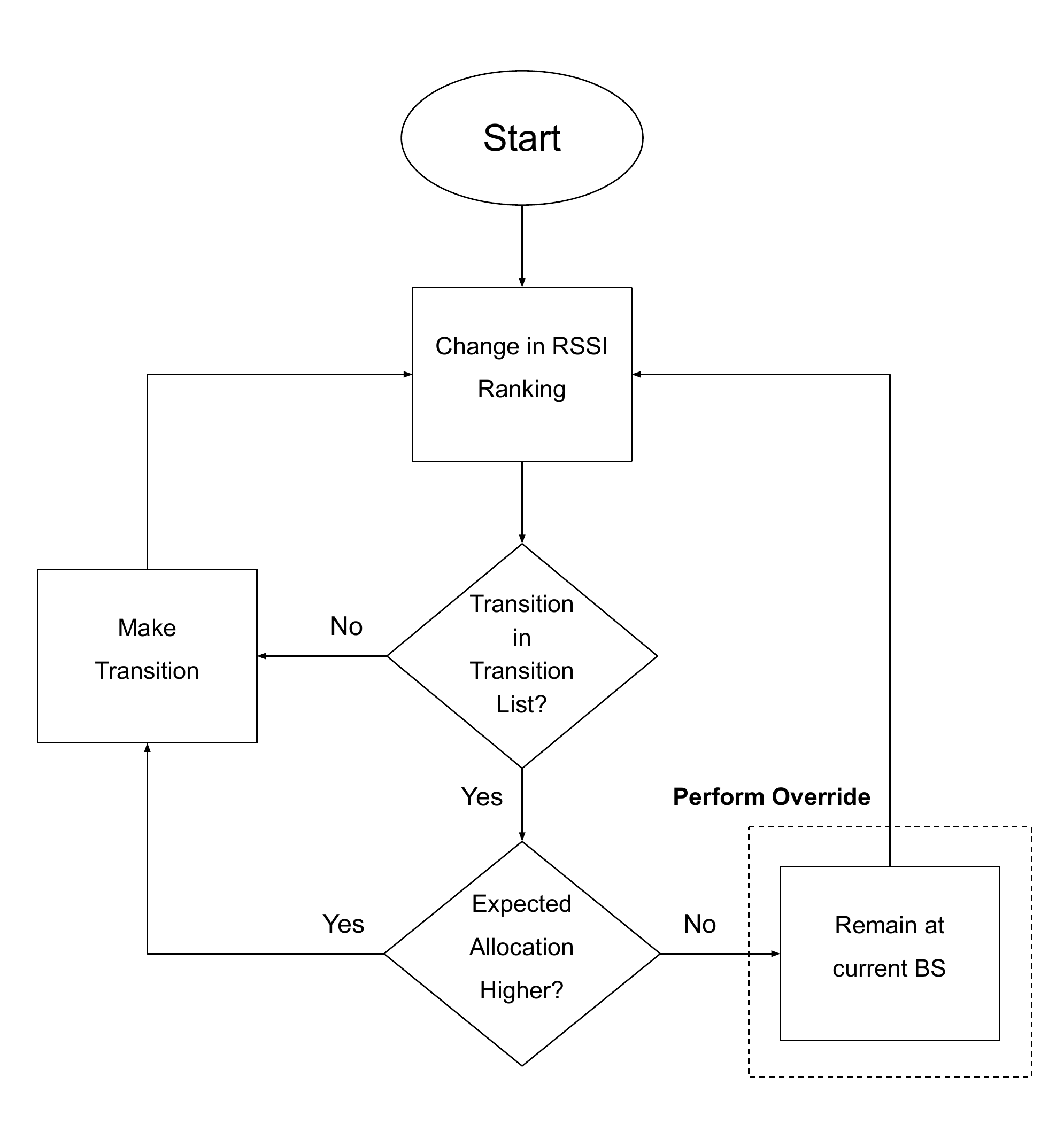}
\caption{Handoff override decision process}
\label{algorithm}
\end{figure} 

Until this point, the UE has been configured with a baseline behavior that mirrors a standard association based on RSSI. To extend this we contribute a new algorithm that learns network allocation outcomes when the rank order of the 3 closest base stations is changed. If the UE has not seen a specific transition before, it continues the default behavior and associates to the base station with the highest RSSI. As the UE performs further handoffs and stores the state transitions, if the UE has seen some $transition_n$ previously, it can choose to perform an "override" and not to perform the handoff if it has learned there is a negative $allocation\ \Delta$ expected in that transition. The decision logic of this override process is shown in figure \ref{algorithm}. The \textit{compute} complexity of the logic is fixed at O(1) due to the logic always using the same two inputs of current allocation and expected allocation to make a decision. The total \textit{algorithmic} complexity of the transition learning process becomes O(log n) when paired with a binary search algorithm, assuming transitions are stored as a sorted list \cite{gcomm-b37}.

\subsubsection{Simulation Environment} 
\begin{figure}
\centering
\includegraphics[width=0.9\linewidth]{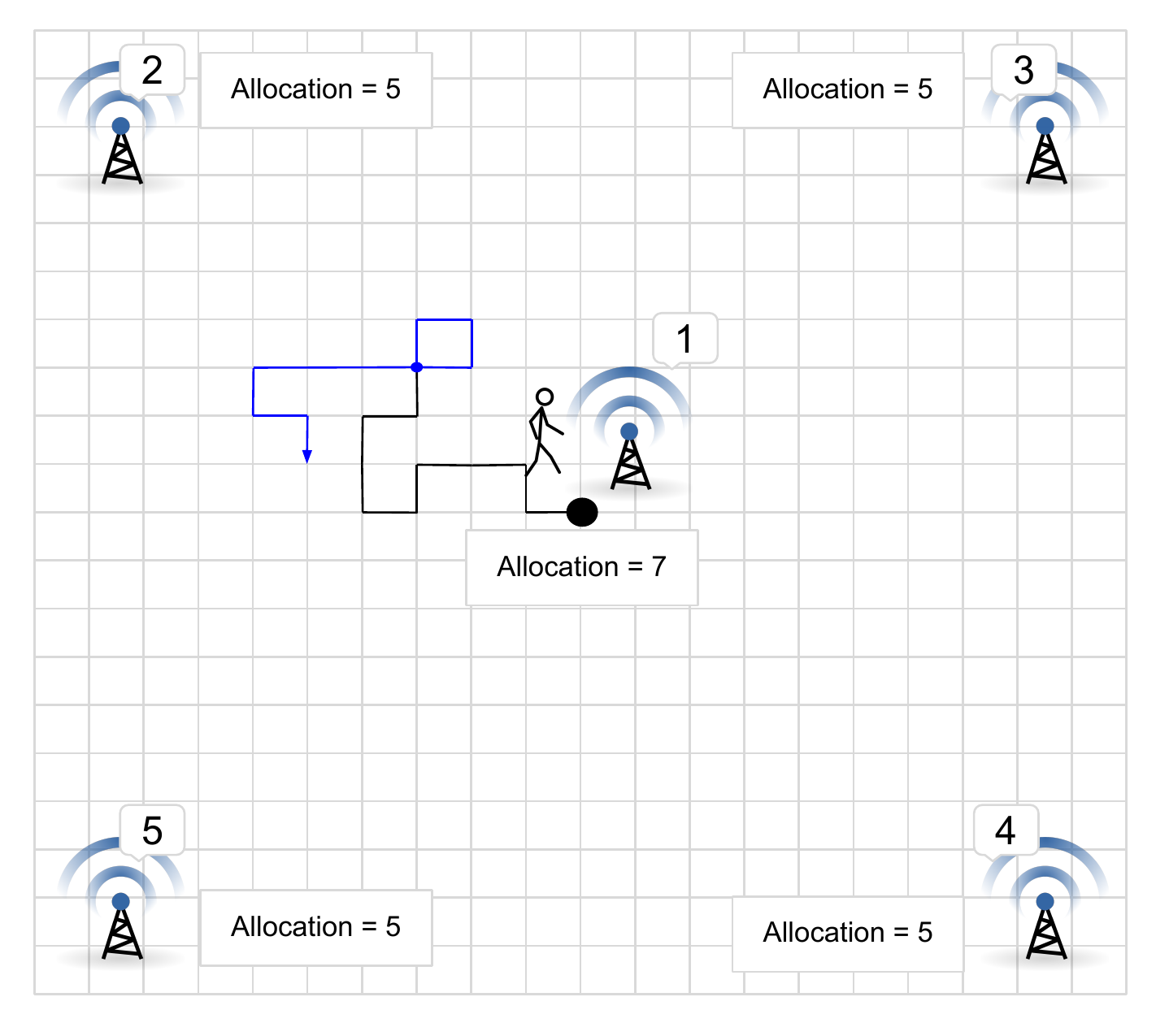}
\caption{The simulation environment using a 10-step random walk and static base station allocations.}
\label{grid_world}
\end{figure}

For the simulation we create an area that is a 23x23 unit grid containing 5 base stations placed at grid positions [0, 0], [22, 0], [22, 22], [22, 0], and [11, 11] (fig. \ref{grid_world}). In this structure the simulation environment presents a 2-D Markov chain with matching state space and cardinality (\ref{prob_matrix}). Each grid unit of the simulation represents 1 city block. 

At the start of the simulation, a UE is placed at position [11, 11] and completes a series of 2000 continuous random walks of 10 unit steps each throughout the environment. Having all grid positions being equidistant and with an eigenvalue of 1, the sum of probability of the UE transitioning into any given position in the state space converges to 1 after the 2000 walk trial (\ref{sum_probability}) \cite{gcomm-b11}. Additionally, setting a boundary for the simulation environment makes the grid state space irreducible, and combining this with the aperiodicity of the random walk enforces that the probability of the UE arriving to any single space in the environment during a single walk is dependant on the point in which the random walk started (\ref{position_probability}) \cite{gcomm-b11}\cite{gcomm-b31}. Transitions and allocations experienced during each 2000 walk trial are then averaged to provide an average allocation result for the simulation round. A total of 1000 simulation such rounds were run in order to provide a monte carlo sample of the transition learning algorithm performance. The simulation environment is written in the Python programming language and is available to download from Github \cite{gcomm-b10}.

\vspace{6pt}
\begin{equation}
P = \begin{bmatrix} 
    P_{0,0} & P_{0,1} & \dots & P_{0,j} & \dots & P_{0,S}\\
    P_{1,0} & P_{1,1} & \dots & P_{1,j} & \dots & P_{1,S}\\
    \vdots & \vdots & \ddots & \vdots & \ddots & \vdots \\
    P_{i,0} & P_{i,1} & \dots & P_{i,j} & \dots & P_{i,S} \\
    \vdots & \vdots & \ddots & \vdots & \ddots & \vdots \\
    P_{S,0} & P_{S,1} & \dots & P_{S,j} & \dots & P_{S,S}
    \end{bmatrix}
    \label{prob_matrix}
\end{equation}
\vspace{6pt}


\begin{equation}
    \mathlarger{\mathlarger{\sum}}_{j=1}^{S}{P_{ij}=1}
    \label{sum_probability}
\end{equation}

\begin{equation}
    \lim_{k\to\infty} (P^k)_{ij} = \pi j
    \label{position_probability}
\end{equation}
\vspace{6pt}

\section{Simulation Results}

\begin{figure}
\centering
\includegraphics[width=\linewidth]{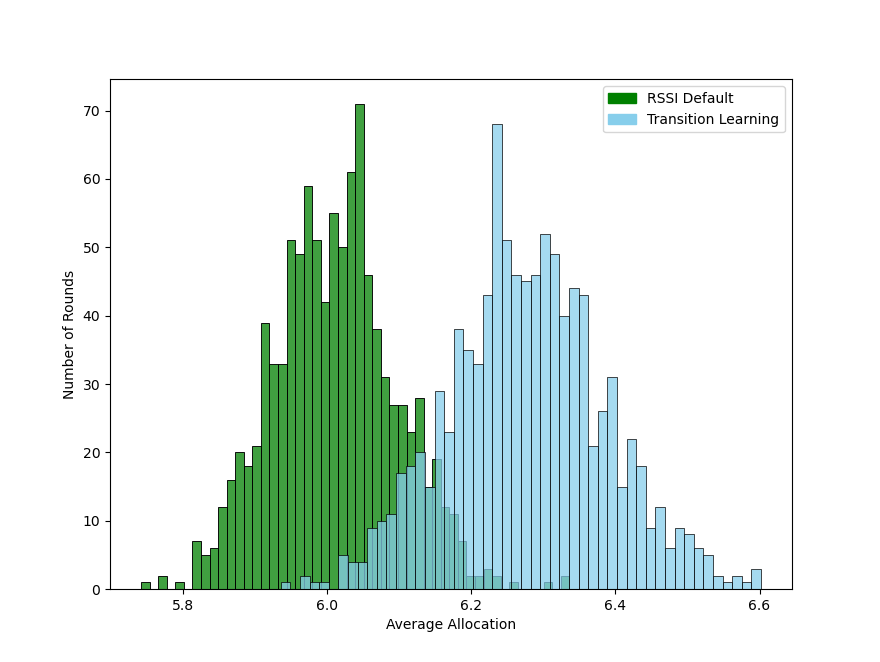}
\caption{Network allocation distribution over 2000 rounds (Default Environment).}
\label{map_4}
\end{figure} 

\begin{figure}
\centering
\includegraphics[width=\linewidth]{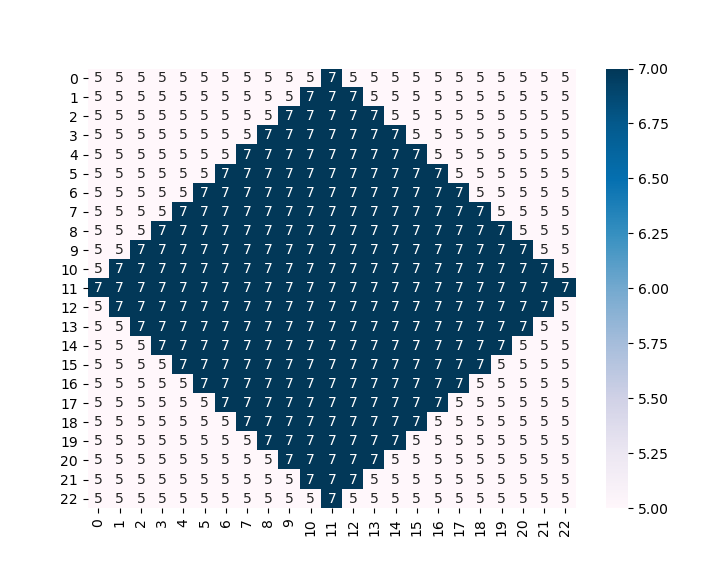}
\caption{Allocation map of Scenario 1 (Default Environment).}
\label{map_1}
\end{figure} 

\begin{figure}
\centering
\includegraphics[width=\linewidth]{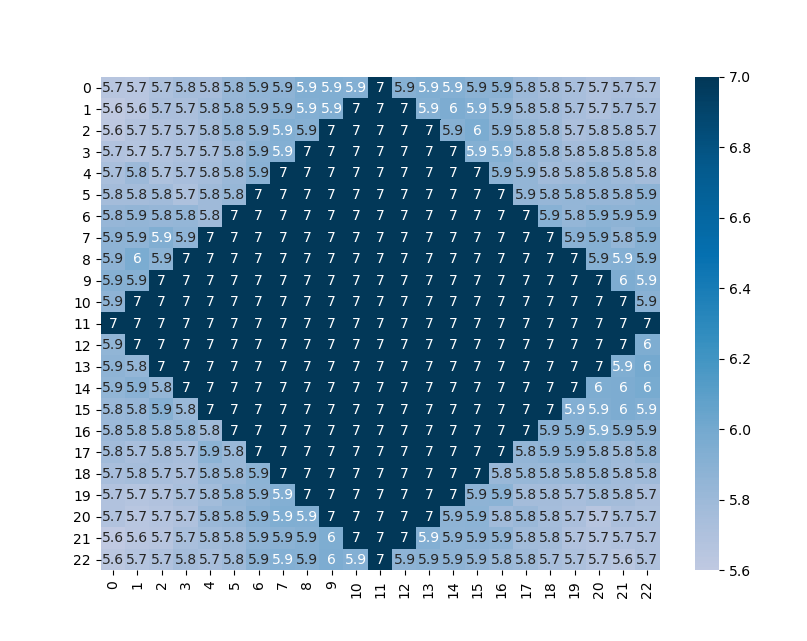}
\caption{Average allocation performance over 2000 rounds using transition learning (Default Environment).}
\label{map_3}
\end{figure} 

\begin{figure}
\centering
\includegraphics[width=\linewidth]{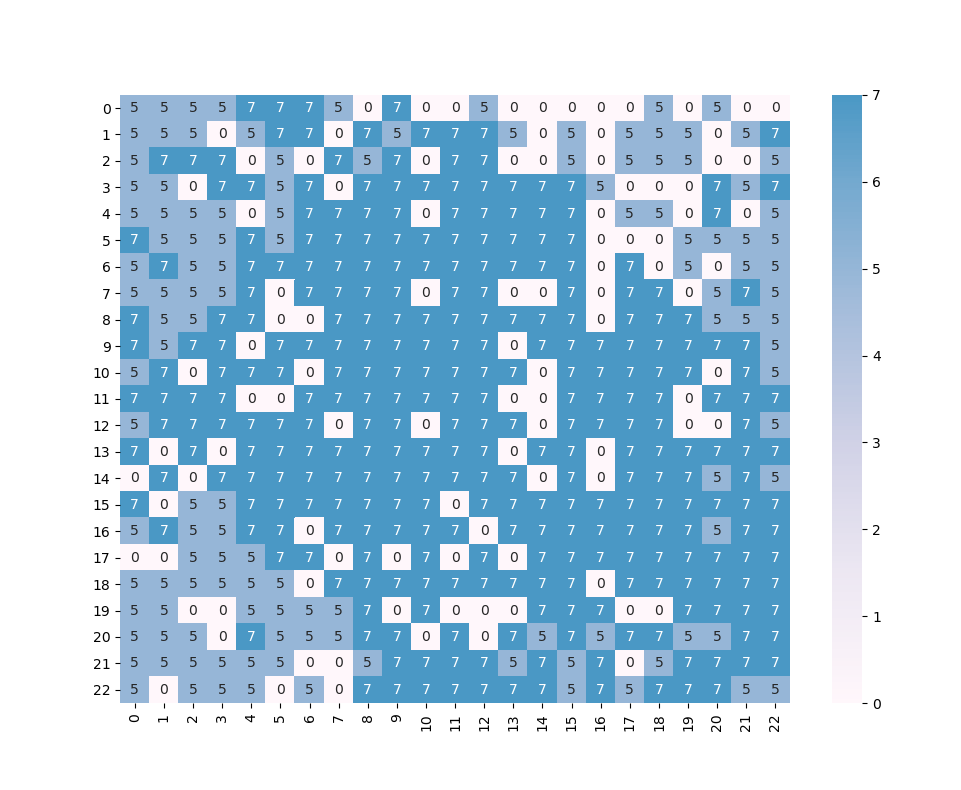}
\caption{Monte Carlo sample of a single random walk round. Allocations marked "0" are unexplored states. (Default Environment).}
\label{map_2}
\end{figure} 


\begin{figure}
\centering
\includegraphics[width=\linewidth]{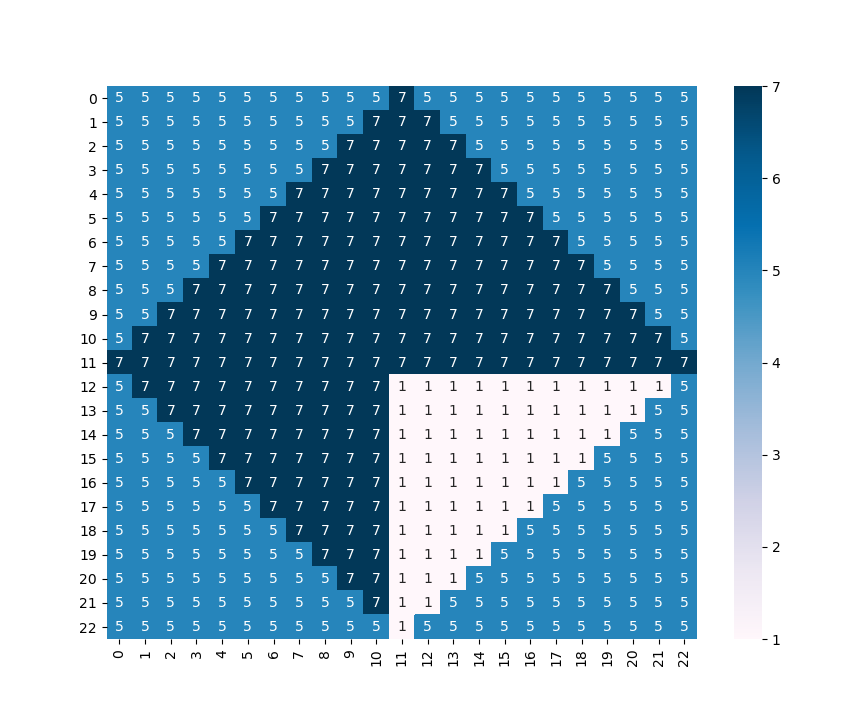}
\caption{Allocation map of Scenario 2 (Sector Load).}
\label{map_5}
\end{figure} 

\begin{figure}
\centering
\includegraphics[width=\linewidth]{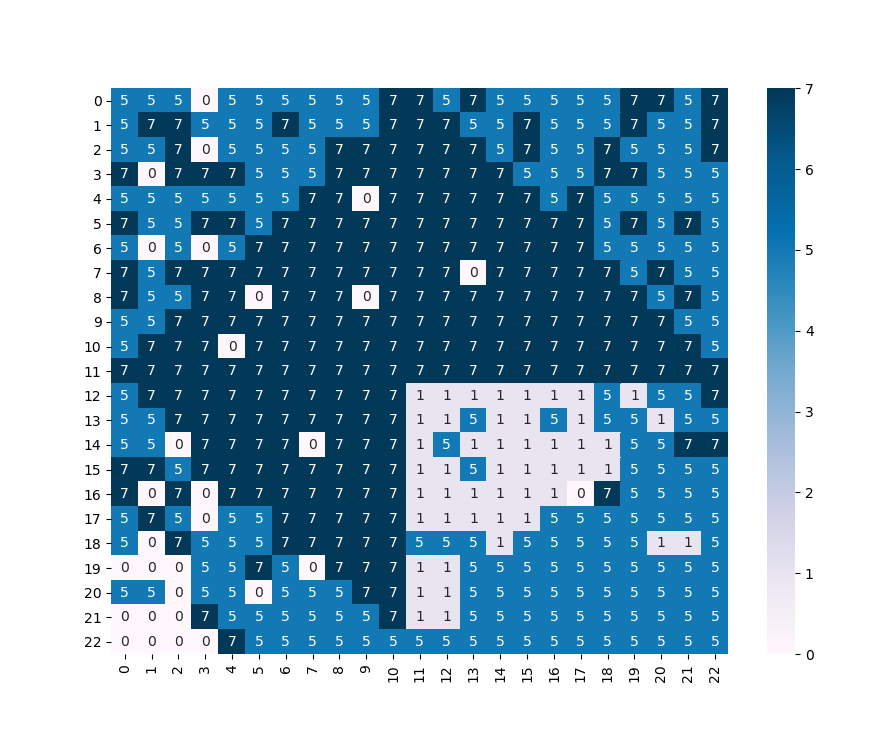}
\caption{Monte Carlo sample of a single random walk round. Allocations marked "0" are unexplored states. (Sector Load).}
\label{map_6}
\end{figure} 

\begin{figure}
\centering
\includegraphics[width=\linewidth]{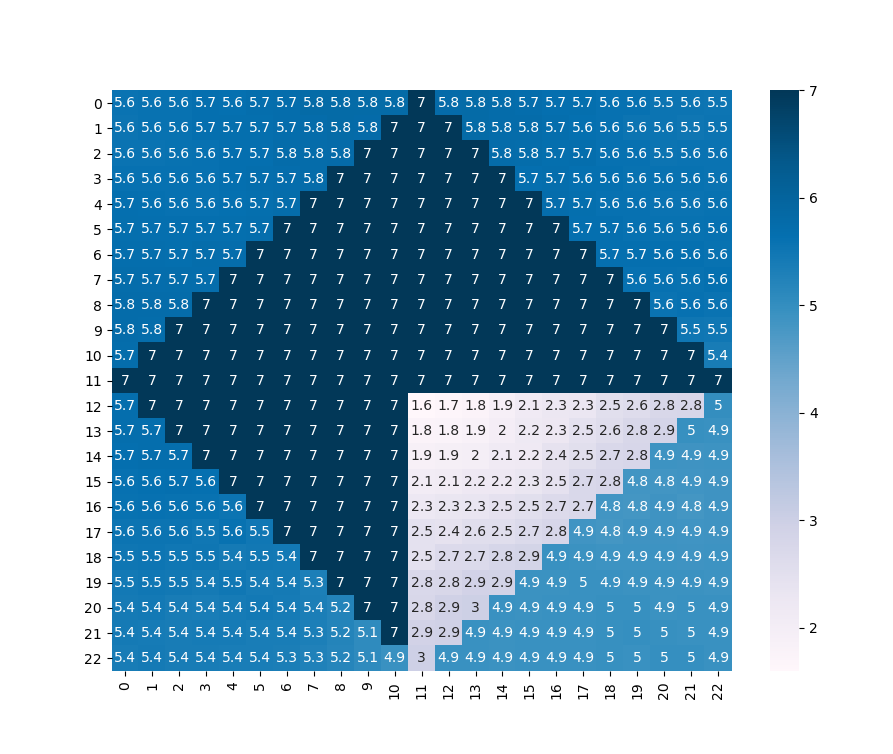}
\caption{Average allocation performance over 2000 rounds using transition learning (Sector Load).}
\label{map_7}
\end{figure} 

\begin{figure}
\centering
\includegraphics[width=\linewidth]{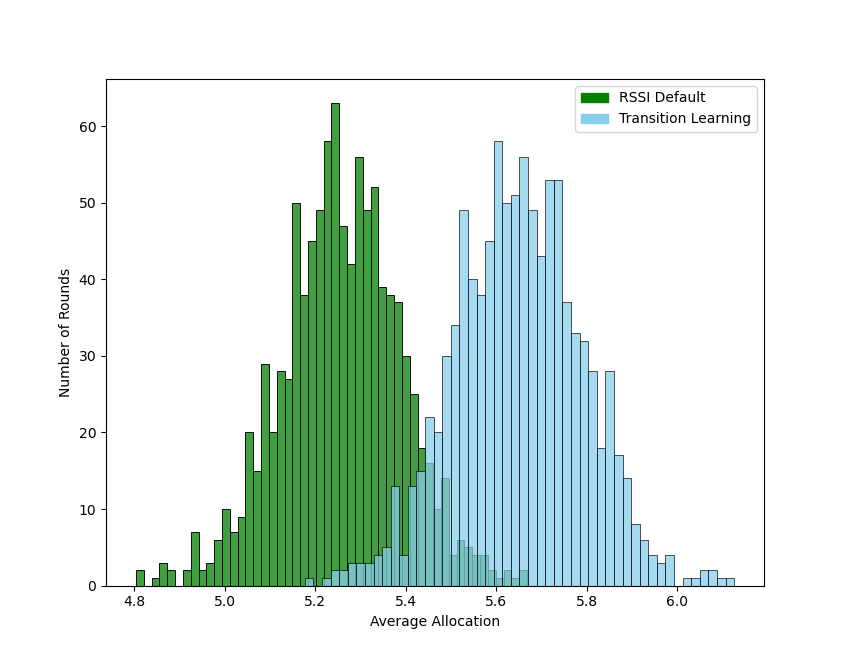}
\caption{Network allocation distribution over 2000 rounds (Sector Load).}
\label{map_8}
\end{figure}

To characterize the performance of the transition learning algorithm we analyse it under 2 scenarios. The results of the two simulation scenarios are presented in table \ref{simulation_table}.

\subsubsection{Scenario 1: Default Environment} The first scenario is the "Default Environment" representing a best case scenario state where the allocations of all base stations is uniform across the entire state space and the final allocation granted is impacted only by the choice of base station association. Within the Default Environment, on average the transition learning algorithm performed an override during 29.36\% of transitions delivering a net allocation increase of 5.5\% compared to base station associations relying only on RSSI (fig. \ref{map_4}). This scenario provided a predictable result where the amount of overrides performed is roughly correlated to the area of the state space occupied by the base station with higher allocation given the environment geometry (fig. \ref{map_1}). This result also affirms the original probability relationship that over 2000 rounds the probability of the UE existing in a given space within the environment becomes 1 (\ref{sum_probability}). Figure \ref{map_7} reveals a pattern of higher average allocation in areas bordering the higher allocation zone, corresponding to the increased probability that a random walk from this area has an increased probability of experiencing a transition or transition override resulting from a base station rank change (\ref{position_probability}). Figure \ref{map_2} provides a single round snapshot that gives a higher resolution example of the overrides and resulting allocations that are occurring during individual rounds.  

\subsubsection{Scenario 2: Sector Load} The second scenario evaluated is "Sector Load" and is representative of a scenario where within the coverage of a single base station, there is some subset of coverage (in this case 1 base station sector) that is under significant load, even while RSSI across the state space is unchanged. In this loaded sector, allocation is changed from \textit{7} to \textit{1} (fig. \ref{map_5}. In this scenario, knowledge of the additional load is not present in the measures available to the UE and is effectively \textit{hidden}. Within the Sector Load scenario, on average the transition learning algorithm performed an override during 30.89\% of transitions delivering a net allocation increase of 7.0\% compared to base station associations based only on RSSI (\ref{map_8}). The amount of overrides performed in this scenario is not significantly changed in this scenario, reflecting the proportion of the state space with an allocation other than \textit{5} remains unchanged. The pattern of increased average allocation near edges of higher allocation is repeated here (fig. \ref{map_7}), but is now shifted towards the base station at position [22,22], reflecting some portion of transitions being learned and then subsequently overridden when involving the base station sector under load. Figure \ref{map_6} again gives a higher resolution example of the overrides and resulting allocations that are occurring during individual rounds of our Scenario 2 with sector load. 


\begin{table*}[htbp]
\centering
\begin{tabular}{l|c|c|c|c|c|c|}
\cline{2-7}
 &
  \multicolumn{3}{c|}{\textbf{Default Environment}} &
  \multicolumn{3}{c|}{\textbf{Sector Load}}\\ \cline{2-7} 
 &
  \cellcolor[HTML]{EFEFEF}\textit{\% Override} &
  \cellcolor[HTML]{EFEFEF}\textit{Allocation Average} &
  \cellcolor[HTML]{EFEFEF}\textit{Performance Gain} &
  \cellcolor[HTML]{EFEFEF}\textit{\% Override} &
  \cellcolor[HTML]{EFEFEF}\textit{Allocation Average} &
  \cellcolor[HTML]{EFEFEF}\textit{Performance Gain} \\ \hline
\multicolumn{1}{|r|}{RSSI Default} &
  0 &
  6.01 &
  - &
  0 &
  5.26 &
  - \\ \hline
\multicolumn{1}{|r|}{Transition Learning} &
  41.59 &
  6.36 &
  5.5\% &
  30.89 &
  5.65 &
  7.0\% \\ \hline
\end{tabular}
\vspace{6pt}
\caption{Summary of simulation results}
\label{simulation_table}
\end{table*}

\section{Discussion and Future Research}

Collectively, the authors present this paper as an early result exploring the broader topic of how a network, or more specifically, UE devices can potentially operate after increases in network decentralization. Being able to place additional environment logic at the UE allows the logic the potential to become agnostic and move \textit{with} the UE in a state where network operation occurs peer-to-peer. It is important to note that of the results achieved, raw performance gain values can be considered as secondary, as they are partially a function of the difference between the chosen allocation values during simulation. The primary experiment finding is the underlying behavior relationships and reliability of the transition learning algorithm to attain a better result with O(log n) complexity - even with hidden environmental contexts such as base station section load. 


A potential area of investigation extending from the presented results is the impact and interaction of having multiple UE's within the environment making mobility decisions based on the transition learning algorithm. In this case it can be assumed that all UE's learn similar outcomes from similar transitions and begin to shift network load. In this case such behavior would bring the problem statement closer to existing reinforcement learning experiments in wireless and allow a further comparison of the two.

\end{document}